\documentclass[a4paper,twoside,reqno]{bjp}
\usepackage{graphicx}
\usepackage{cite}
\usepackage{amssymb,amsmath,amscd,amsthm}
\usepackage{times}
\usepackage{rotating}
\usepackage[bookmarks=false]{hyperref}
\hypersetup{%
    colorlinks=true,        
    linkcolor=blue,          
    citecolor=blue,         
    urlcolor=blue           
    }

\usepackage{geometry}
\allowdisplaybreaks

\begin{document}

\title{A $\gamma$-rigid solution of the Bohr Hamiltonian with deformation-dependent mass term for Kratzer potential and $\gamma$ = 30°}

\runningheads{List of Authors with Abbreviated  Names}{A. Lahbas et al.}

\begin{start}{%
\coauthor{S. Ait El Korchi}{1},
\coauthor{S. Baid}{1},
\coauthor{P. Buganu}{2},
\coauthor{M. Chabab}{1},
\coauthor{A. El Batoul}{1},
\author{A. Lahbas}{3,1},
\coauthor{M. Oulne}{1},

\address{High Energy Physics and Astrophysics Laboratory, Faculty of Sciences Semlalia, Cadi Ayyad University, P. O. B. 2390, Marrakesh 40000, Morocco}{1}
\address{Horia Hulubei$-$National Institute of Physics and Nuclear Engineering, 077125, Bucharest-Magurele, Romania}{2}
\address{ESMaR, Department of Physics, Faculty of Science, Mohammed V University in Rabat,  Rabat 10000, Morocco}{3}

}

\begin{Abstract}
In this work,  the Davydov-Chaban Hamiltonian, describing the collective motion of $\gamma$-rigid atomic nuclei, is amended by allowing the mass parameter to depend on the nuclear deformation. Further, Z(4)-DDM (Deformation-Dependent Mass) model is proposed by considering the Kratzer potential for the $\beta$ variable, and solving the problem by techniques of asymptotic iteration method (AIM). The results of the calculated spectra and $B(E2)$ transition rates for series of $^{192-196}$Pt isotopes are compared with the corresponding experimental data as well as with other theoretical models. Exact analytical expressions are derived for spectra and normalized wave functions of the Kratzer potential. The obtained results show an overall good agreement with the experimental data and an important improvement in respect to other models. \end{Abstract}

\begin{KEY}
Davydov-Chaban Hamiltonian, Shape phase transition, Collective models, Deformation dependent effective mass, Kratzer potential. \end{KEY}
\end{start}

\section{Introduction }
The study of shape phase transitions in nuclear structure have attracted a lot of attention  from both experimental and theoretical perspectives. Therefore, several approaches have been developed in this context especially in the framework of the Bohr-Mottelson Model (BMM) \cite{BMM} and the Interacting Boson Model (IBM)\cite{IBM}.  Furthermore, the interest devoted for this topic has increased even more with the occurrence of Critical Point Symmetries (CPSs). Such symmetries, for example, E(5) \cite{E5} and X(5) \cite{X5} correspond to the shape phase transitions U(5)$\leftrightarrow$O(6) and U(5)$\leftrightarrow$ SU(3), respectively. 
In this context, considerable attempts have been done for several potentials to achieve analytical solutions of the Bohr Hamiltonian, either in the usual case where the mass parameter is assumed to be a constant \cite{buganu2015analytical,budaca2016extended,Raduta:2013db,Lahbas,Levai}, or by introducing the Deformation-Dependent Mass Formalism (DDMF) \cite{bonat10,bonat11,bonat13,chabab2016nuclear,Adahchour}. In addition, another direction of research was to investigate such phenomena by imposing a $\gamma$-rigidity as in the case of
Z(4) \cite{Bonatsos2005} or X(3) \cite{bonatsos2006x}.

In the present work, the attention is focused on the study of the quadrupole collective states in $\gamma$-rigid case, by modifying Davydov-Chaban Hamiltonian \cite{dav60} in the framework of DDMF \cite{quesne2004,CLO15P,Buganu18} with Kratzer  potential \cite{kra20} for the variable $\beta$ and $\gamma$ fixed to $\pi/6$ . The expressions for the  energy levels as well as for the wave functions  are obtained in closed analytical form by means of the Asymptotic Iteration Method (AIM)\cite{AIM03}, an efficient technique that we have used to solve many similar problems \cite{CHABAB:2012qd,clo5,clo6,celo5}. 

The content of this paper is arranged as follows. In section 2, the theoretical framework of Z(4)-DDM model is briefly described. In section 3, the exact separation of the Hamiltonian and solution of angular equation are achieved. The analytical expressions for the energy levels of Kratzer potential and the wave functions are given in Section 4. Finally, Section 5 is devoted to the numerical calculations for energy spectra, $B(E2)$ transition probabilities,  while Section 6 contains the main conclusions. 

\section{Theoretical framework of Z(4)-DDM model}

In the framework of Davydov$-$Chaban model \cite{dav60}, the nucleus is assumed  to be  $\gamma$$-$rigid. Therefore, the Hamiltonian operator depends on four variables $(\beta,\theta_i)$ and has the following form \cite{dav60}:
\begin{equation}
    H=-\frac{\hbar^2}{2B}\left[\frac{1}{\beta^{3}}\frac{\partial}{\partial\beta} {\beta^3}\frac{\partial}{\partial\beta}-
   \frac{1}{4\beta^2}\sum_ {k=1,2,3}\frac{Q_{k}^{2}}{\sin^2(\gamma-\frac{2}{3}\pi k)}\right]
  +V(\beta),
  \label{eq6}
\end{equation}
where $B$ is the mass parameter, $\beta$ the collective coordinate and $\gamma$ a parameter, while $Q_{k}$ are the components of the total angular momentum in the intrinsic frame and $\theta_i$ the Euler angles.

In order to construct the Davydov-Chaban Hamiltonian with a mass depending on the deformation coordinate $\beta$, one follows the formalism described in Sec. II of \cite{bonat11} considering:
\begin{equation}
B(\beta)=\frac{B_0}{f(\beta)^2},
\label{eq7}
\end{equation}
where $B_0$ is a constant and $f(\beta)$ is the deformation function  depending only on the radial coordinate $\beta$. Then, only the $\beta$ part of the resulting equation will be affected by deformation dependent mass.  The resulting equation reads as:

\begin{multline}
 \left[ -\frac{\sqrt{f}}{\beta^3}\frac{\partial}{\partial\beta} {\beta^3f}\frac{\partial}{\partial\beta}\sqrt{f}+
  \frac{f^2}{4\beta^2}\sum_ {k=1,2,3}\frac{Q_{k}^{2}}{\sin^2(\gamma-\frac{2}{3}\pi k)} \right]\Psi(\beta,\Omega)\\+v_{eff}\Psi(\beta,\Omega)=\epsilon\Psi(\beta,\Omega),  \label{eq8}
\end{multline}
with
\begin{equation}
v_{eff}=v(\beta)+  \frac{1}{4}(1-\delta-\lambda)f\bigtriangledown^2f
+\frac{1}{2}\left(\frac{1}{2}-\delta\right)\left(\frac{1}{2}-\lambda \right)(\bigtriangledown f)^{2},
    \label{eq9}
\end{equation}
 where the reduced energies and  potentials are defined as $\epsilon=\frac{B_0}{\hbar^2}E$ and $v(\beta)=\frac{B_0}{\hbar^2}V(\beta)$, respectively.

Considering a total wave function of the form  $\Psi(\beta,\Omega)=\chi(\beta)\phi(\Omega)$, where $\Omega$ denotes the rotation Euler angles ($\theta_1$,$\theta_2$,$\theta_3$), the separation of variables gives  two equations
  \begin{multline}
$\Bigg [$-\frac{1}{2}\frac{\sqrt{f}}{\beta^3}\frac{\partial}{\partial\beta} {\beta^3f}\frac{\partial}{\partial\beta}\sqrt{f}+ \frac{f^2}{2\beta^2}\Lambda +  \frac{1}{4}(1-\delta-\lambda)f\bigtriangledown^2f
$\Bigg]$\chi(\beta)\\+\frac{1}{2}$\Bigg [$$\Big($\frac{1}{2}-\delta$\Big)$$\Big($\frac{1}{2}-\lambda$\Big)$(\bigtriangledown f)^{2}+ v(\beta) $\Bigg]$\chi(\beta)=\epsilon\chi(\beta),  \label{eq10}
\end{multline}
\begin{equation}
\left[\frac{1}{4}\sum_ {k=1,2,3}\frac{Q_{k}^{2}}{\sin^2(\gamma-\frac{2}{3}\pi k)}-\Lambda\right]\phi(\Omega)=0,
    \label{eq11}
\end{equation}
where $\Lambda$ is  the eigenvalue for the equation of the angular part. In the case of $\gamma=\pi/6$, the angular momentum term can be written as \cite{Mey75},
\begin{equation}
\sum_ {k=1,2,3}\frac{Q_{k}^{2}}{\sin^2(\gamma-\frac{2}{3}\pi k)}= 4(Q^2_1+Q^2_2+Q^2_3)-3Q^2_1. \label{eq12}
\end{equation}
Eq. \eqref{eq11} has been solved by Meyer-ter-Vehn \cite{Mey75}, with the results
\begin{equation}
\Lambda=L(L+1)-\frac{3}{4}\alpha^2, \label{eq13}
\end{equation}
\begin{equation}
\phi(\Omega)=\phi^L_{\mu,\alpha}(\Omega)=\sqrt{\frac{2L+1}{16\pi^2(1+\delta_{\alpha,0})}}\left[ \mathcal{D}^{(L)}_{\mu,\alpha}(\Omega) +(-1)^L\mathcal{D}^{(L)}_{\mu,-\alpha}(\Omega) \right], \label{eq14}
\end{equation}
where $\mathcal{D}(\Omega)$ denotes Wigner functions of the Euler angles, $L$ is the total angular momentum quantum number, $\mu$ and $\alpha$   are the quantum numbers of the
projections of angular momentum on the laboratory fixed $z$-axis and the body-fixed $x'$-axis, respectively. In the literature, for the triaxial shapes, it is customary to insert the wobbling quantum number $n_w$ instead of $\alpha$, with $n_w=L-\alpha$ \cite{Mey75}. Within this convention, the eigenvalues of the angular part become :
\begin{equation}
\Lambda=L(L+1)-\frac{3}{4}(L-n_w)^2. \label{eq15}
\end{equation}

\section{Z(4)-DDM solution for $\beta$ part of the Hamiltonian}
The $\beta$-vibrational states of the triaxial nuclei, having a $\gamma$ rigidity of $\pi/6$, are determined by the solution of the radial Schr\"odinger equation
\begin{multline}
\frac{1}{2}f^2\chi'' +\left(\frac{3f^2}{2\beta}+ff'\right)\chi' +\left(\frac{3ff'}{4\beta}+\frac{(f'^2)}{8}+\frac{ff''}{4}   \right)\chi\\-\frac{f^2}{2\beta^2}\Lambda\chi+\epsilon\chi-v_{eff}\chi=0,
 \label{eq16}
\end{multline}
with
\begin{equation}
v_{eff}=v(\beta)+  \frac{1}{4}(1-\delta-\lambda)ff''
+\frac{1}{2}\left(\frac{1}{2}-\delta\right)\left(\frac{1}{2}-\lambda \right)( f')^{2}.
    \label{eq17}
\end{equation}
By using the standard transformation of the radial wave function as $\chi(\beta)=\beta^{-3/2}R(\beta)$, one gets:
\begin{equation}
f^2R''+2ff'R'+(2\epsilon-2u_{eff})R=0,
    \label{eq18}
\end{equation}
with
\begin{equation}
u_{eff}=v_{eff}+\frac{f^2}{2\beta^2}\Lambda+\left(\frac{3ff'}{4\beta}+\frac{3f^2}{8\beta^2}-\frac{(f')^2}{8}-\frac{ff''}{4} \right).
    \label{eq19}
\end{equation}
 Now, one considers the special case of the Kratzer potential type \cite{kra20}, defined as 
\begin{equation}
v(\beta)=-\frac{1}{\beta}+\frac{\beta_0}{2\beta^2},
    \label{eq20}
\end{equation}
where $\beta_0$ indicates the position of the minimum of the potential. According to the specific form of the potential \eqref{eq20}, one chooses the deformation function in the following special form \cite{bonat13}:
\begin{equation}
f(\beta)=1+a\beta, \hspace{1.5cm}  a<<1.    \label{eq21}
\end{equation}
By inserting  the potential and the deformation function in Eq. \eqref{eq18}, one gets:
\begin{equation}
2u_{eff}(\beta)=k_0+\frac{k_{-1}}{\beta}+\frac{k_{-2}}{\beta^2},   \label{eq22}
\end{equation}
with
\begin{align}
k_{0\  }=&\frac{a^2}{2}\Big[4+2\Lambda+2(\frac{1}{2}-\delta)(\frac{1}{2}-\lambda)+ 3(1-\delta-\lambda)\Big],  \nonumber&
\nonumber\\
k_{-1}=&a[3+2\Lambda+\frac{3}{2}(1-\lambda-\delta)]-2,   \nonumber&
\nonumber\\
k_{-2}=&\frac{3}{4}+\Lambda+\beta_0. &
  \label{eq23}
\end{align}
In order to apply the asymptotic iteration method  \cite{AIM03}, we propose an appropriate physical wave function for the radial function $R(\beta)$:
\begin{equation}
R_{n_{\beta}L}(\beta)=\beta^{\eta}(1+a\beta)^{\kappa}G_{n_{\beta}L}(\beta),
\label{eq24}
\end{equation}
with
\begin{align}
\eta=&\frac{1}{2}(1+\sqrt{1+4k_{-2}}),  \nonumber&
\nonumber\\
\kappa=& -\frac{1}{2}\Bigg(1+\sqrt{1+4k_{-2}+\frac{4k_0}{a^2}-\frac{8\epsilon}{a^2}-\frac{4k_{-1}}{a}} \Bigg).&
  \label{eq25}
\end{align}
For this form,  the radial wave equation reads:
\begin{align}
G''(\beta)=&-\Bigg[\frac{2(\eta+a\beta(1+\eta+\kappa))}{\beta(1+a\beta)}\Bigg]G'(\beta)\\-&\Bigg[ \frac{2a\eta(\eta+\kappa)-k_{-1}}{\beta(1+a\beta)}\Bigg]G(\beta),
\label{eq26}
\end{align}
while the generalized formula of the radial energy spectrum is:
\begin{equation}
\epsilon_{n_{\beta}n_{w}L}=\frac{1}{2}\left[k_0+\frac{a^2}{4}-\left( \frac{k_{-1}+an_{\beta}^2+a\eta(1+2n_{\beta})}{2(\eta+n_{\beta})}  \right)^2 \right],
\label{eq27}
\end{equation}
where $n_{\beta}$ is the principal quantum number of $\beta$ vibrations.

The excited-state wave functions  reads: 
   \begin{multline}
R_{n_{\beta}L}(\beta)=C_{n_{\beta}L}\ a^{-\eta} \ 2^{n_{\beta}+\kappa_n}\ (1+t)^{\eta} \ (1-t)^{-\kappa_n -\eta-n_{\beta}}  \\\times P_{n_{\beta}}^{(-1-2(n_{\beta}+\kappa_{n}+\eta),2\eta-1)}(t), \label{eq40}
\end{multline}
 with $t=\frac{-1+a\beta}{1+a\beta}$, while 
\begin{align}
C_{n_{\beta}L}=&\left[\frac{a^{1+2\eta}}{2}\ \frac{(1+2n_{\beta}+2\eta+2\kappa_n)(1+2\kappa_n)}{(\eta+n_{\beta})} \right]^{\frac{1}{2}}\nonumber \\ \times & \left[\frac{\Gamma(1+n_{\beta}+2\kappa_n)\Gamma(n_{\beta}+1)}{\Gamma(n_{\beta}+2\eta+2\kappa_n)\Gamma(n_{\beta}+2\eta)}     \right]^{\frac{1}{2}}.
\label{eq42}
\end{align}
The quantities $k_0$, $k_{-1}$, $k_{-2}$ are given by Eq. \eqref{eq23}, while $\Lambda$ is the eigenvalue of the angular part given by Eq. \eqref{eq15}. The excitation energies depend on three quantum numbers, $n_{\beta}$, $n_{w}$ and $L$, respectively four parameters: $a$ the deformation mass parameter, $\beta_0$ the minimum of the potential,  the free parameters $\delta$ and $\lambda$ coming from DDM formalism \cite{roos83}. In the last part of the paper, a comparison to the experiment will be carried out by fitting the theoretical spectra to the experimental data. Finally, it will be shown that the predicted energy levels turn out to be independent of the choice made for $\delta$ and $\lambda$.
\section{Numerical results }
The Z(4)-DDM model, presented in the previous sections, has been applied for calculating the energies of the collective states and the reduced $E2$ transition probabilities for $^{192,194,196}$Pt isotopes. All bands ($i.e.$ ground state, $\beta$ and $\gamma$) are characterized by three quantum numbers $n_{\beta}$, $n_w$ and $L$.  The ground state band (g.s.) is characterized by $n_{\beta}=n_w=0$, the $\beta$-band by $n_{\beta}=1, n_w=0$, and the $\gamma$-band by $n_{\beta}=0, n_w=2$ for even $L$ levels and $n_{\beta}=0, n_w=1$ for odd $L$ levels.

In this work, the theoretical predictions for the levels \eqref{eq27}, are treated equally, depending on two parameters, namely: the potential minimum $\beta_0$ and the deformation dependent mass parameter $a$. These parameters are adjusted to reproduce the experimental data by applying a least-squares fitting procedure for each considered isotope. We evaluate the root mean square (rms) deviation between the theoretical values and
the experimental data via the formula: 
\begin{equation}\label{eq54}
\sigma=\sqrt{\frac{\sum_{i=1}^m(E_i(Exp)-E_i(th))^2}{(m-1)E(2_g^+)}}.
\end{equation}
\setlength{\tabcolsep}{4pt}
\begin{sidewaystable}
{
\caption{The energy spectra comprising the ground, $\gamma$ and $\beta$ bands obtained with our models Z(4)-DDM Kratzer (K) are compared with the values taken from \cite{buganu2015analytical} and \cite{budaca2016extended} with the available experimental data \cite{A192,A194,A196}.\\\label{table:table1}}
\begin{tabular}{ccccccccccccccccccccc}
\hline\noalign{\smallskip}
 &\multicolumn {4}{c}{ $^{192}$Pt} & &\multicolumn {4}{c}{ $^{194}$Pt}&&\multicolumn {4}{c}{ $^{196}$Pt}\\
\cline {2 -5}\cline {7 -10}\cline {12 -15}
 &Exp \cite{A192}&K&Ref.\cite{buganu2015analytical}&Ref.\cite{budaca2016extended}&&Exp \cite{A194}&K&Ref.\cite{buganu2015analytical}&Ref.\cite{budaca2016extended}&&Exp \cite{A196}&K&Ref.\cite{buganu2015analytical}&Ref.\cite{budaca2016extended}\\
\noalign{\smallskip}\hline\noalign{\smallskip}

R$_{0,0,4}$ &2.479 & 2.451&2.439 & 2.396&$$ & 2.470&2.506&2.415&2.406&$$&2.465&2.455&2.513&2.481\\
R$_{0,0,6}$&4.314  & 4.129&3.787 & 3.834&$$ & 4.298&4.334&3.835&3.902&$$&4.290&4.144&3.709&3.701\\
R$_{0,0,8}$&6.377  & 5.844&5.773 & 5.761&$$ & 6.392&6.306&5.880&5.896&$$&6.333&5.877&5.579&5.559\\
R$_{0,0,10}$&8.624  & 7.472&7.350 & 7.484&$$ &8.672&8.280&7.573&7.713&$$&8.558&7.528&6.914&6.932\\
\noalign{\bigskip}
R$_{1,0,0}$ &3.776 & 3.768&3.397 & 3.537&$$ & 3.858&3.806&3.706&3.809&$$ &3.192&3.124&2.954&2.977\\
R$_{1,0,2}$&4.547 & 4.472&4.995 & 5.162&$$ &4.603&4.555&5.409&5.493&$$ &3.828&3.844&4.308&4.364\\
R$_{1,0,4}$&  &5.506 &7.002 & 7.113&$$ & $$&7.511&5.693&7.490&$$ &&4.904&6.238&6.280\\
\noalign{\bigskip}
R$_{0,2,2}$&1.935 & 1.900&1.653 & 1.664&$$ & 1.894&1.926&1.661&1.676&$$ &1.936&1.902&1.646&1.643\\
R$_{0,1,3}$&2.910  & 2.714&2.302 & 2.345&$$ & 2.809&2.786&2.332&2.378&$$ &2.852&2.719&2.249&2.252\\
R$_{0,2,4}$&3.795  & 4.548&4.229 & 4.200&$$ & 3.743&4.806&4.268&4.273&$$ &3.636&4.566&4.179&4.150\\
R$_{0,1,5}$&4.682  & 4.748&4.342 & 4.360&$$ & 4.563&5.034&4.402&4.446&$$ &4.526&4.769&4.243&4.227\\
R$_{0,2,6}$&5.905  & 6.785&6.358 & 6.466&$$ & $$&7.434&6.524&6.645&$$ &5.644&6.830&6.041&6.049\\
R$_{0,1,7}$&6.677  & 6.637&6.065& 6.215&$$ & $$&7.255&6.235&6.392&$$ &&6.681&5.737&5.754\\
R$_{0,2,8}$&8.186  & 8.640&9.163 & 9.203&$$ & $$&9.764&-&9.508&$$ &7.730&8.717&8.564&8.573\\
\noalign{\smallskip}\hline\noalign{\smallskip}
rms&  &0.500 & 0.614 &0.593 &$$ &&0.390&0.543&0.515&&&0.576&0.682&0.683\\
a& & 0.002  &  & &$$ &&0.004&&&&&0.006&&\\
$\beta_0$& & 50.0  &  & &$$ &&70.6&&&&&51.0&&\\
\noalign{\smallskip}\hline

\end{tabular}
}
\end{sidewaystable}

\setlength{\tabcolsep}{6.8pt}
\begin{sidewaystable}
\caption{The comparison of experimental data \cite{A192,A194,A196} (upper line) for several $B(E2)$ ratios of  nuclei to predictions by the Davydov-Chaban Hamiltonian with $\beta$-dependent mass for the Kratzer potential (lower line), using the parameter values shown in Table \ref{table:table1}. \\\label{table:table2}}
{
\begin{tabular}{lllllllllll}
\hline
 nuleus & $\frac{4_g\rightarrow 2_g}{2_g\rightarrow 0_g}$ & $\frac{6_g\rightarrow 4_g}{2_g\rightarrow 0_g}$ &$\frac{8_g\rightarrow 6_g}{2_g\rightarrow 0_g}$ & $\frac{10_g\rightarrow 8_g}{2_g\rightarrow 0_g}$ & $\frac{2_{\gamma}\rightarrow 2_g}{2_{1}\rightarrow 0_g}$&$\frac{2_{\gamma}\rightarrow 0_g}{2_g\rightarrow 0_g}$&$\frac{0_{\beta}\rightarrow 2_g}{2_g\rightarrow 0_g}$&$\frac{2_{\beta}\rightarrow 0_g}{2_g\rightarrow 0_g}$\\
 &&&&&&$\times 10^3$&&$\times 10^3$ & \\
\hline

  $   ^{192}$Pt & $1.56(12)$ & $1.23(55)$&$ $ & $$&$1.91(16)$ & $9.5(9)$&$$&$$&\\
  & $1.58$ & $2.38$&$3.32$ & $4.61$&$1.60$ & $0.0$&$0.70$&$17.36$&0.2966\\\\

   $   ^{194}$Pt & $1.73(13)$ & $1.36(45)$&$1.02(30) $ & $0.69$&$1.81(25)$ & $5.9(9)$&$0.01$&$$&\\
  & $1.56$ & $2.31$&$3.19 $ & $4.41$&$1.58$ & $0.0$&$0.67$&$25.08$&0.6239\\\\

    $   ^{196}$Pt & $1.48(3)$ & $1.80(23)$&$1.92(23) $ & $$&$$ & $0.4$&$0.07(4)$&$0.06(6)$&\\
  & $1.63$ & $2.58$&$3.86$ & $5.91$&$1.65$ & $0.0$&$0.90$&$23.16$&0.3751\\
\hline
\end{tabular}
}
\end{sidewaystable}

This quantity, represents the $rms$ deviations of the theoretical calculations from the experiment, where $m$ denotes the number of states, while $E_i(exp)$ and $E_i(th)$ represent the theoretical and experimental energies of the i-th level, respectively. $E(2_g^+)$ is the energy of the first excited level of the ground state  band (gsb). 

From Table \ref{table:table1}, one can see that the obtained results for the levels belonging to gs, $\beta$ and $\gamma$ bands are in quite satisfactory agreement with experimental data. Analyzing the mean deviation  corresponding for each nucleus, one can see that the present results are fairly better that those obtained by Z(4)-sextic model. This is explained by the fact that here the mass parameter depends on the $\beta$ variable, while in Refs. \cite{buganu2015analytical,budaca2016extended} the mass is considered as a constant.

Similarly, we have calculated the intraband and interband $B(E2)$ transition rates, normalized to the transition from the first excited level of the ground state band to the ground state, using the same optimal values  of the three parameters obtained from fitting the energy ratios. From the obtained theoretical results shown in Table \ref{table:table2}, one can remark some discrepancies within the ground state band of the higher $L$ levels, while the experimental values show a decreasing trend.  For the interband transitions rates from the $\gamma$ band to the gsb, our model give good results, while interband transitions from the $\beta$ band to the gsb, the agreement is only partially good.

\section{Conclusion }
In this paper, based on Davydov-Chaban Hamiltonian within the framework of DDM formalism, a new model entitled  Z(4)-DDM has been elaborated. The numerical realization of this model consists in calculating the energy spectra and transition probabilities of $^{192,194,196}$Pt isotopes using Kratzer as collective potential compared to experimental data as well as with other models results. In addition, we have shown that the present model has well improved predictions in comparison with those of \cite{buganu2015analytical,budaca2016extended}.

\section*{Acknowledgments}
A. Lahbas would like to thank the organizing committee for the wonderful scientific meeting.

\end{document}